\pdfoutput=1
\documentclass[twocolumn, prl,amsmath,amssymb,floatfix,superscriptaddress]{revtex4-1}
\usepackage[latin9]{inputenc}
\usepackage{amsmath}
\usepackage{amssymb}
\usepackage{graphicx}
\usepackage{esint}

\makeatletter
\@ifundefined{textcolor}{}
{%
 \definecolor{BLACK}{gray}{0}
 \definecolor{WHITE}{gray}{1}
 \definecolor{RED}{rgb}{1,0,0}
 \definecolor{GREEN}{rgb}{0,1,0}
 \definecolor{BLUE}{rgb}{0,0,1}
 \definecolor{CYAN}{cmyk}{1,0,0,0}
 \definecolor{MAGENTA}{cmyk}{0,1,0,0}
 \definecolor{YELLOW}{cmyk}{0,0,1,0}
}

\usepackage{bm}
\@ifundefined{definecolor}
 {\usepackage{color}}{}
\usepackage{epsfig}\usepackage{subfigure}

\newcommand{\be}{\begin{equation}}
\newcommand{\ee}{\end{equation}}

\makeatother

\begin{document}

\title{ Quantum diffusion in the strong tunneling regime}

\author{Nisarga Paul}

\affiliation{Department of Physics, Harvard University, Cambridge, Massachusetts
02138, USA}
\author{ Ariel Amir}
\affiliation{School of Engineering and Applied Sciences, Harvard University, Cambridge, Massachusetts 02138, USA}

\begin{abstract}
We study the spread of a quantum-mechanical wavepacket in a noisy
environment, modeled using a tight-binding Hamiltonian. Despite the
coherent dynamics, the fluctuating environment may give rise to diffusive
behavior. When correlations between different level-crossing events
can be neglected, we use the solution of the Landau-Zener problem
to find how the diffusion constant depends on the noise. We also show that when an electric field or external disordered potential is applied to the system, the diffusion constant is suppressed with no drift term arising. The results are relevant to various quantum systems, including exciton diffusion in photosynthesis and electronic transport in solid-state physics. 
\end{abstract}
\maketitle

\section{Introduction}

As Anderson showed more than half a century ago, a quantum particle
in a one-dimensional disordered potential does not diffuse, but rather
stays confined in a finite region of space, its wavefunction being
exponentially localized \cite{Anderson}. Since, this problem has
been extensively studied, theoretically, numerically and experimentally
\cite{abrahams}. When the disorder fluctuates in time, however, completely
different behavior emerges: for a particle in a continuous, time-dependent
potential, this leads, remarkably, to super-diffusive behavior \cite{superdiffusion,rosenbluth,fishman},
as was experimentally demonstrated \cite{segev}. For particles
on a lattice, diffusive behavior has been shown to arise, both for noise which is
uncorrelated in time \cite{diffusion_jetp,post}, and more recently
for noise with a finite correlation time $\tau$ \cite{amir_PRE}.
The latter study has shown that the diffusion constant depends on
the correlation time and strength of the noise in a non-trivial way,
with a ``motional-narrowing'' regime at small $\tau$, and a correlation-time-independent
diffusion constant at sufficiently large $\tau$. These results,
however, were only applicable for the case of small tunneling, allowing
the utilization of the separation of scales between the fast dephasing
process and the slow diffusion. Here, we extend the results to large
tunneling matrix elements, relying on the exact solution of the Landau-Zener
problem. This approach gives an intuitive understanding of the emergence of diffusion from the quantum dynamics. We also show that when an electric field or external disorder is applied in addition to the noisy environment, it does not lead to any drift but on the contrary leads to a diminished diffusion constant. Our model may be viewed as a simplified version of models used to study exciton diffusion in photosynthesis \cite{photosynthesis1,photosynthesis2}.

\section{Model and derivation}

 We study the dynamics of a single particle
on a lattice, described by the Schr\"odinger equation, where the
on-site energies are randomly fluctuating. The time-dependent Hamiltonian is given by

\begin{equation}
H=\sum_{j}T(c_{j}^{\dagger}c_{j+1}+c_{j+1}^{\dagger}c_{j})+x_{j}(t)c_{j}^{\dagger}c_{j},\label{eq:hamiltonian}
\end{equation}

with the noise term $x_j(t)$ assumed to be Gaussian and uncorrelated
between different sites, but correlated in time according to $\langle x_i(t')x_j(t'+t)\rangle=\delta_{ij}C(t)$.
The typical time
of the decay of $C(t)$ is defined as the correlation time $\tau$, the typical magnitude is defined as $W$ so that $C(0)=W^{2}$, and the typical noise velocity is defined $v= c W/\tau$ with $c$ a dimensionless constant. 

In the absence of an electric field, it was shown previously \cite{amir_PRE}
that for weak tunneling (compared to all other energy scales in the
problem) and for arbitrary correlation functions of the (Gaussian)
noisy environment, a classical master equation for diffusion arises, and the diffusion constant can be found analytically. We now extend these results to the case of strong tunneling using a different approach.

\subsection{Landau-Zener approach} 
A beautiful, exactly solvable problem
regards the transition probability of a quantum particle in a two-level
system \cite{landau_zener1,landau_zener2}.
Under the assumptions that the spacing between the levels, $\Delta E,$
is ramped linearly in time, and that that at time $t=-\infty$ the
particle is known to be in one of the levels, the transition probability
is given by:
\begin{equation}
p=1-e^{-2\pi T^{2}/v_C},\label{eq:LZ}
\end{equation}

where $T$ is the tunneling constant, $v_C$ is the crossing velocity (i.e., $\frac{d\Delta E}{dt}=v_C$)
and we set $\hbar=1$.

\begin{figure}
\includegraphics[width=4.25cm]{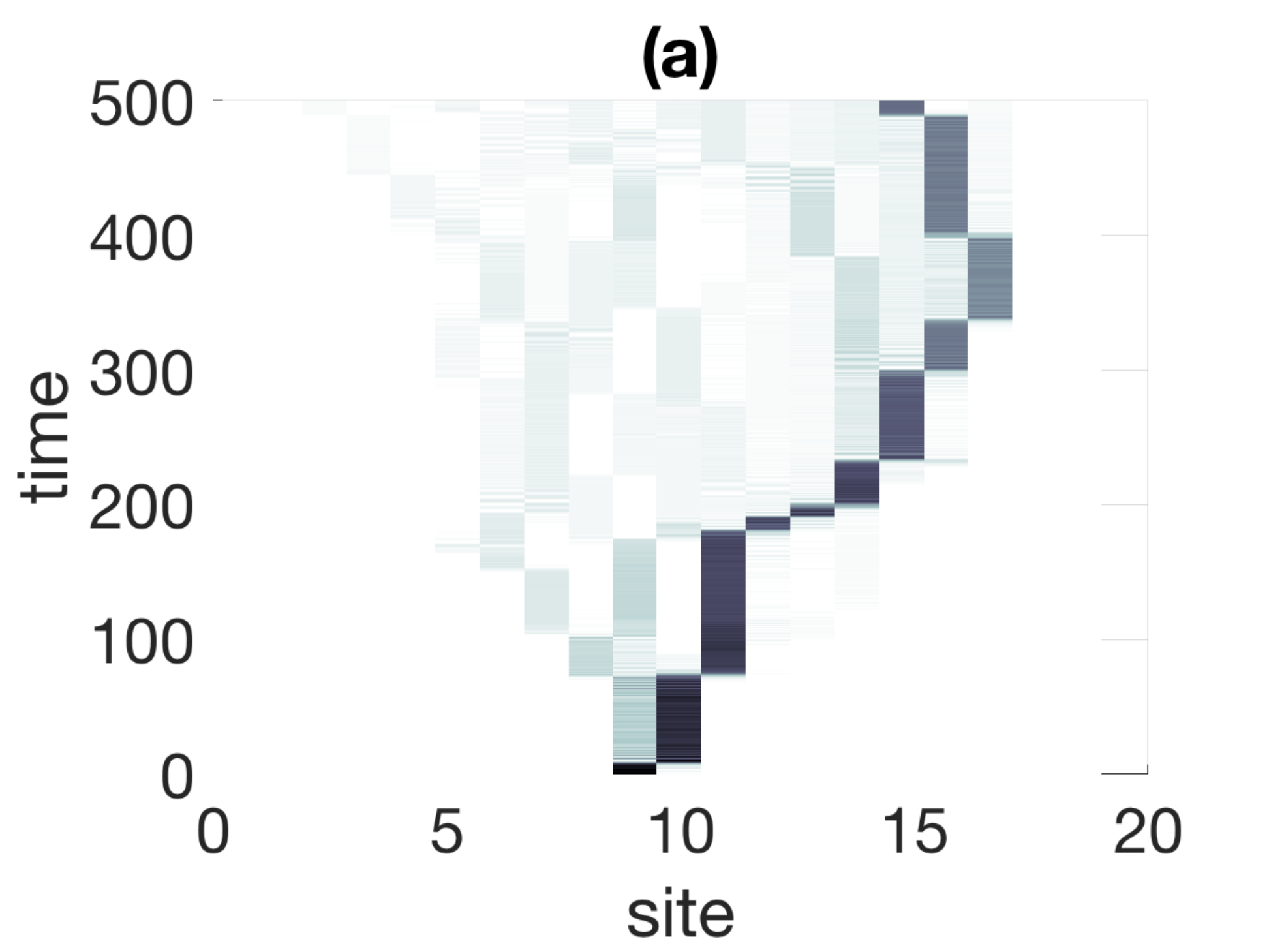}
\includegraphics[width=4.25cm]{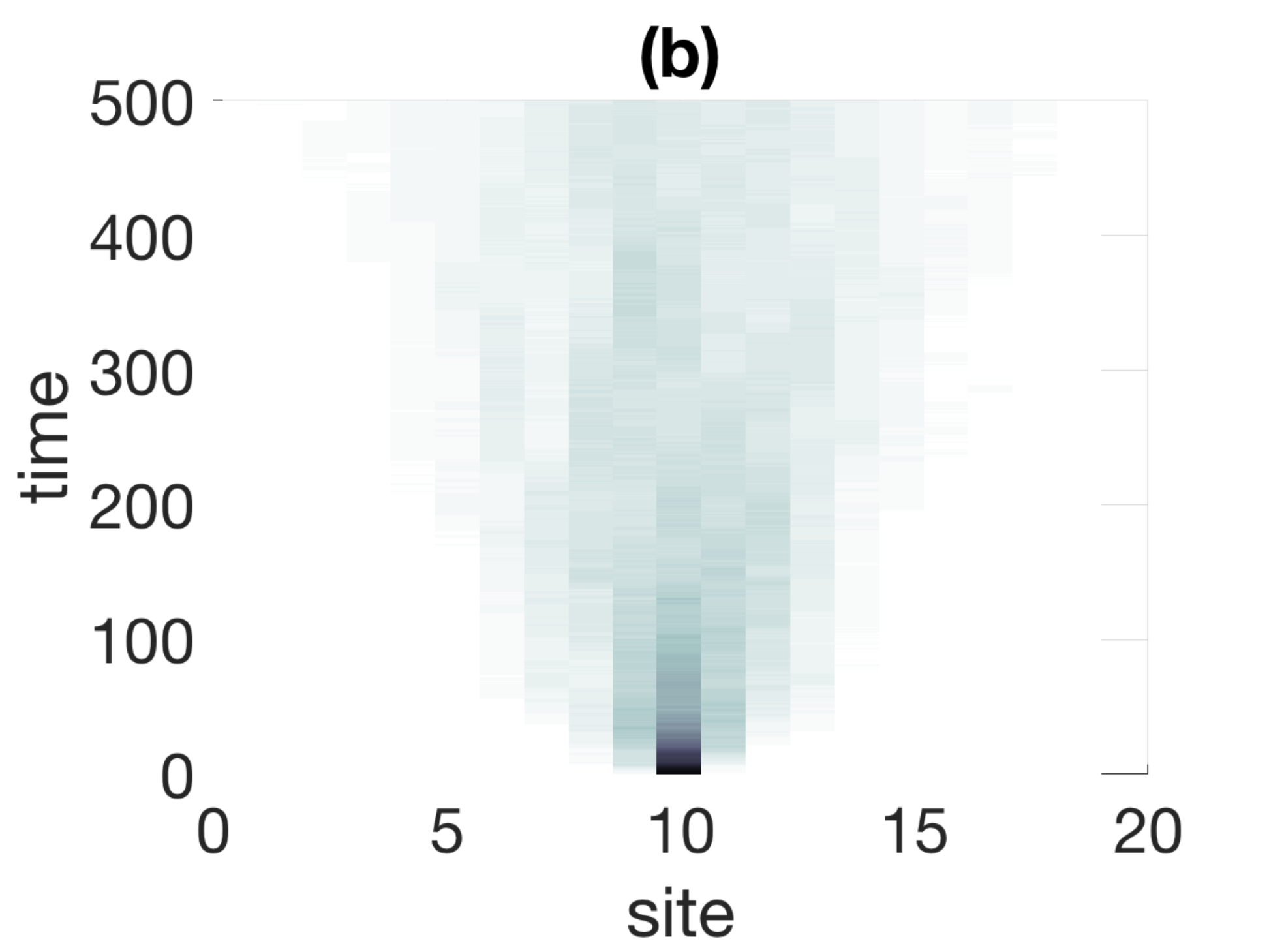}
\caption{Time evolution of amplitude profile of an initially single-site wavepacket  (a) for one trial and (b) averaged over 50 trials ($W=10,T=0.5,c=1,\tau=50$). The points in (a) where the amplitude peak jumps coincide with level crossings. The noise is generated from Eq. \eqref{eq:sde} and a fourth-order Runge-Kutta solver was used to solve the Schr\"odinger equation.} \label{fig:diff}
\end{figure}

We may now think of the original noisy Hamiltonian as driving multiple
transitions, by making neighboring levels cross again and again. Since
the dynamics is random, it is clear that one should obtain classical
diffusion in this way if the probabilities of hopping between sites
at different crossing events are independent. In Fig. \ref{fig:diff}, this diffusion is illustrated qualitatively by simulating the time evolution under Eq. \eqref{eq:hamiltonian} of a wavepacket. \par

If the transition probability at each level crossing can be described by Eq. \eqref{eq:LZ} and $f$ is the frequency of level crossings, the resulting diffusion constant is 

\be
D= f \langle1-e^{-2\pi T^{2}/v_C}\rangle, \label{lz_res}
\ee
which involves an ensemble average over $v_C$, the velocities of crossing events. In order for the distribution of $v_C$ to be well-defined, the noise samples must be taken to be differentiable. We consider noise samples $x(t)$ defined by the second-order stochastic differential equation

\be
m \ddot x + \eta \dot x + k x = \xi(t), \label{eq:sde}
\ee

 a harmonic oscillator equation with mass $m$,  damping parameter $\eta$, and spring constant $k$, driven by Gaussian white noise $\xi(t)$ with $\langle \xi(t)\xi(t')\rangle = \delta(t-t')$. As detailed in Appendix \ref{app:noise}, the parameter choices

\begin{equation}\label{eq:subs}
m = \frac{\tau^{3/2}}{W c\, \widetilde c},\quad \eta = \frac{\tau^{1/2} \widetilde c}{2W c},\quad k = \frac{c}{W\tau^{1/2} \widetilde c},
\end{equation}

with $\widetilde c \equiv \sqrt{2(c^2+1)}$ and $c\geq 1$ provide a noise with correlation time $\tau$ and stationary distribution

\be\label{eq:P(X,V)}
P(x,v) = \frac{\tau}{2\pi cW^2} \exp\left( -\frac{x^2}{2W^2} -  \frac{v^2}{2c^2W^2/\tau^2} \right)
\ee
where $v = \dot x$. We remark that first-order noise such as that generated by an Ornstein-Uhlenbeck process would fail to have a well-defined stationary distribution for $v$, as can be seen from Eqs. \eqref{eq:subs},\eqref{eq:P(X,V)} by fixing $W,\tau$ and taking $m\to 0$ (via $c\to \infty)$. \par

The statistics of $v_C$ is equivalent to the statistics of $\frac{d}{dt} \left. y\right|_{y=0}$ for a sample $y$ with twice the variance of $P(x,v)$, since the difference of two noise samples also obeys Eq. \eqref{eq:sde} albeit with noise of twice the variance. Denoting this distribution $\widetilde P$, the distribution of crossing velocities is proportional to $\widetilde P(0, v) |v|$ because a random walker traversing an interval $[-\Delta x, \Delta x]$ with velocity $v$ will be weighted in the distribution $\widetilde P(0,v)$ by the time spent in the interval. Therefore the distribution of crossing velocities is
\begin{equation}\label{eq:dist}
 \rho(v_C) = \frac{\exp\left( - \frac{v_C^2}{4c^2W^2/\tau^2}\right)}{4c^2W^2/\tau^2} |v_C|.
\end{equation}

 To be able to use the Landau-Zener (LZ) probability and subsequently Eq. \eqref{lz_res}, we must satisfy three conditions: (i) the finite duration of the crossing event should be sufficiently long (as Eq. \eqref{eq:LZ} is only exact at $t=\infty$), (ii) the probability of neighboring levels interfering with LZ transitions must be negligible, and (iii) the crossing should be well approximated to be \textit{linear} in time. \par
For condition (i), we note that the duration $t_{LZ}$ of an LZ transition scales as

\begin{equation}\label{eq:tLZ}
t_{LZ} \sim \max\left(  \frac{1}{\sqrt{v_C}} , \frac{T}{v_C}\right),
\end{equation} 

which is derived in Ref. \cite{mullen} using an ``internal clock'' approach which we summarize in Appendix \ref{app:LZ}. 
For Eq. \eqref{eq:LZ} to be applicable to the crossings, the frequency $f$ of crossing events must therefore satisfy $f^{-1} \gg t_{LZ}$. We will later see that $f = \frac{c}{\pi \tau}$, and from Eq. \eqref{eq:dist} we find $v_C \sim cW/\tau$ to be the typical crossing velocity.  From Eq. \eqref{eq:tLZ}, it follows that for $T^2 \lesssim cW/\tau$ we obtain the constraint $W\tau \gg c$, and for $T^2 \gtrsim cW/\tau$ we obtain $T\ll W$. 

For condition (ii), we note that an LZ two-level crossing occurs in a window of size $\sim v_C t_{LZ}$. During the crossing, an additional neighboring site energy traverses a window of size $\sim v t_{LZ}$ where $v\sim v_C$. Since the additional level is distributed as a Gaussian of width $W$ (Eq. \eqref{eq:P(X,V)}), the probability that it crosses the LZ crossing will be negligible provided $v t_{LZ} \ll W$. The constraints that follow are identical to those of condition (i).

\par

For condition (iii), we require $\frac{\Delta v}{v} \ll 1$, where $\Delta v$ is the change in relative velocity between two crossing noise samples over time $t_{LZ}$. To obtain an expression for $\Delta v$, we take the difference of two solutions of Eq. \eqref{eq:sde} and integrate over a time $t_{LZ}$, noting the $x$ term vanishes at a crossing:

\be
m \Delta v \approx \eta v t_{LZ} + \sqrt{2t_{LZ}}. 
\ee
The typical size of $v$ at a crossing is $2cW/\tau$ and $\eta,m$ are given by Eq. \eqref{eq:subs}. For $t_{LZ}\sim \frac{1}{\sqrt{v}}$, we have 

\be
\frac{\Delta v}{v} \sim \frac{\widetilde c(2^{3/4}(W\tau c)^{1/4}+\widetilde c)}{2^{3/2}(W\tau c)^{1/2}}
\ee

and for $t_{LZ} \sim \frac{T}{v}$ we have 

\be
\frac{\Delta v}{v} \sim \frac{\widetilde c(2(WTc)^{1/2}+\widetilde cT)}{4cW}.
\ee 
The constraints obtained from $\frac{\Delta v}{v} \ll 1$ are 

\begin{equation}\label{eq:conditions}
\frac{W}{T} \gg 1,c,\qquad W\tau \gg 1, c^3,
\end{equation}
which subsume the constraints from conditions (i), (ii) and, in particular, do not restrict us from the strong-tunneling regime $T\gg \tau^{-1}.$\par

\begin{figure}
\center
\includegraphics[width=8.7cm]{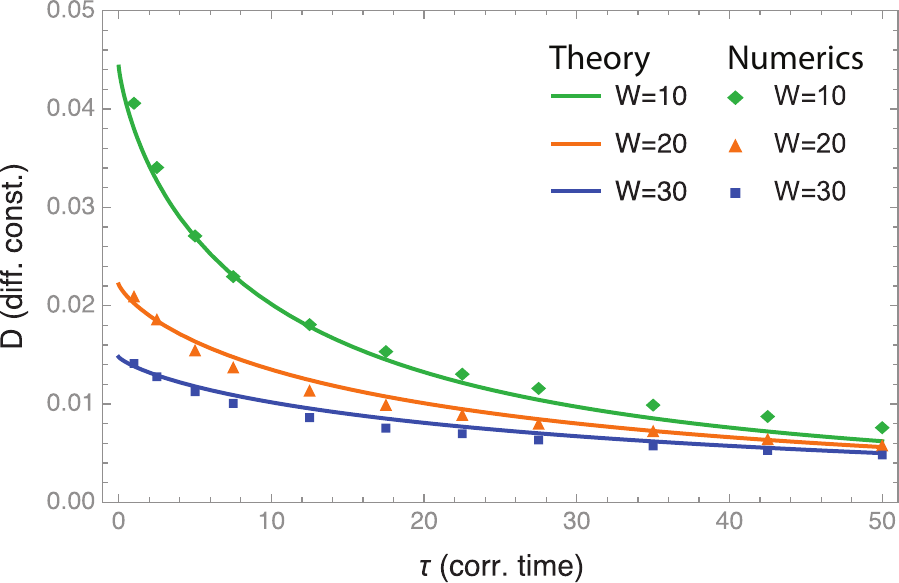}
\caption{Comparison of Eq. \eqref{eq:formula} and numerical integration of the Schr\"odinger equation is shown for the range $0 < \tau < 50$ and model parameters $T=0.5,c=1,$ and $W=10,20, 30$. Each point represents an average over $10^3$ runs of a fourth-order Runge-Kutta solver for the Schr\"odinger equation. The implicit midpoint method from Ref. \cite{burrage} has been used to generate the noise. No fitting parameters are used in the plot.}
\label{fig:plot1}
\end{figure}

Now we proceed to find a closed-form expression for the diffusion constant. With the observation that $f = \frac{c}{\pi \tau}$ (Appendix \ref{app:rice}), we may evaluate Eq. (\ref{lz_res}):

\begin{align}
D &= \frac{c}{\pi\tau} \int_{-\infty}^{\infty} \left( 1- e^{ - 2\pi T^2/|v_C|}\right) \rho(v_C) dv_C\nonumber\\
\label{eq:formula}
&= \frac{c}{\pi\tau}-\frac{\pi ^{1/2} \tau  T^4
   G_{0,3}^{3,0}\left(\frac{\pi ^2 T^4 \tau ^2}{4 c^2 W^2}|
\begin{array}{c}
 -1,-\frac{1}{2},0 \\
\end{array}
\right)}{ 4c W^2}
\end{align}
 where $G^{m,n}_{p,q}$ is the Meijer-G function. In Fig. \ref{fig:plot1}, we compare this formula with simulations of time evolution under the Hamiltonian and find good agreement, with no fitting parameters. As we discuss below, while $f$ and $\rho(v_C)$ are exact, Eq. \eqref{eq:formula} is an approximation since it ignores correlations between subsequent crossing events.  
 
  \par 
We may use the small $x$ expansion

 \begin{equation}
 G_{0,3}^{3,0}\left(x|
\begin{array}{c}
 -1,-\frac{1}{2},0 \\
\end{array}
\right) = \frac{\sqrt{\pi}}{x} - \frac{2\pi}{\sqrt{x}}  + O(\ln x)
 \end{equation}
 
to observe that  Eq.  \eqref{eq:formula} reduces to $D = \frac{\sqrt{\pi} T^2}{W}$  in the weak tunneling limit $W\tau \gg 1,c^3$ and $T^2 \ll v$, in agreement with the analytical results from Ref. \cite{amir_PRE}. In the opposite limit, the Landau-Zener probability approaches unity so that the particle hops at every crossing and $D \to \frac{c}{\pi\tau}$. Eq. \eqref{eq:formula} provides an exact interpolation between these two asymptotic limits, a regime which was inaccessible in previous works. We remark that the reasoning used here is independent of spatial dimension; while the existence of a mobility edge in Anderson localization only occurs above two dimension, in the case of a fluctuating potential diffusive behavior will occur in any dimension. 
\par

 \begin{figure}
\center
\includegraphics[clip,width=8.7cm]{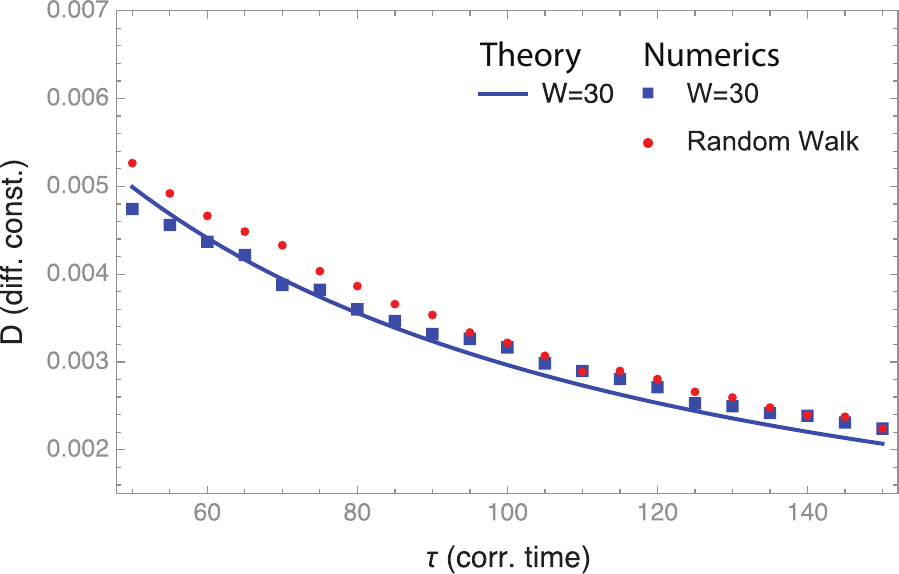}
\caption{Diffusion constants of random walkers which hop with the LZ probability at crossings between numerically generated noise samples are plotted (red circles) alongside theory (blue line) and quantum simulations (blue squares) for $W=30,T=0.5,c=1,$ and $50<\tau<150$. The random walkers are correlated as detailed in the main text. We observe that the quantum simulations, generated as in Fig. \ref{fig:plot1}, agree with the classical random walk simulations for large $\tau$, the regime of validity of the LZ approach (see Eq. \eqref{eq:conditions}).}
\label{fig:plot2}
\end{figure}

In Fig. \ref{fig:plot1} we observe small but systematic deviations from theory for large $\tau$. Accounting for this is a classical, stochastic effect: for large $\tau$, the LZ probability is $\approx 1$ and the particle can be described as a classical random walker, hopping at every crossing. If we denote the hopping directions by $d_n = \pm 1$, we find numerically that $\langle d_n d_{n+1} \rangle > 0$, so that the classical random walker is correlated. This holds persistently for large $\tau$ and originates from Eq. \eqref{eq:sde} having inertia. For large $\tau$ and $T\ll W$, we find a deviation of $\sim$ 7\%. In Fig. \ref{fig:plot2}, this effect is shown to suitably account for the deviations.

\subsection{Addition of an electric field or external disorder}
\label{sec:E}

A priori one might expect that adding an electric field to the Hamiltonian will generate in addition to the diffusion a drift or Bloch oscillations. However, we show analytically that this is not the case, and that in the presence of an electric field, at long times the dynamics is diffusive, with a \textit{suppressed} diffusion constant and no drift. 
\par
If the site energies are shifted $x_j(t) \mapsto x_j(t) - jE$, the statistics of crossing velocities for adjacent sites is equivalent to the statistics of $\left.\frac{d}{dt}y\right|_{y=0}$ for a sample $y$ with twice the variance of $P(x,v)$ and mean $\langle x\rangle = E$. Denoting this distribution $\widetilde P_E$, the distribution of crossing velocities is proportional to $\widetilde P_E(0,v)|v|$ by the same argument which precedes Eq. \eqref{eq:dist}.  Normalizing, we obtain the same distribution as Eq. \eqref{eq:dist}. While the statistics of crossing velocities remains the same, the crossings become less frequent. Indeed, from Appendix \ref{app:rice} we find that the frequency becomes $f = \frac{c}{\pi \tau}e^{- \frac{E^2}{4W^2}}$, and as a result the electric field suppresses the diffusion constant by a factor $e^{- \frac{E^2}{4W^2}}$.
\par For an intuitive explanation of this phenomenon, we note that a noisy bath is equivalent to a thermal bath at infinite temperature. Considering the Einstein relation $D = \sigma k_B T$, where $\sigma$ is conductance, we find that a finite diffusion constant at infinite temperature implies a negligible conductance. This explains why the electric field does not lead to a finite drift, which would be obtained, for example, when the particle is coupled to a finite-temperature phonon bath \cite{mahan}.\par
The above analysis would still hold when an electric field is replaced by an external disorder. Since only $\langle x_j - x_{j+1}\rangle = E$ is used in the analysis, the results would be essentially the same when replacing $E$ by the magnitude of the external disorder. In Appendix \ref{app:E}, the suppression of the diffusion constant by $e^{- \frac{E^2}{4W^2}}$ is reproduced using the independent methods of Ref. \cite{amir_PRE}, applicable for weak tunneling.

\section{Conclusion}

We have considered the dynamics of a single particle in the tight-binding model with random time-dependent on-site energies. We have modeled the diffusion as arising from the energy level crossings between neighboring sites, approximated as Landau-Zener crossings. This approach allows us to extend to the regime of strong tunneling, which was inaccessible in previous works. Upon choosing a model for the random noise with well-defined velocity, we have found a closed-form expression for the diffusion constant which agrees well with simulations. We have further showed that the addition of an electric field or external disordered potential suppresses the diffusion constant, and one does not obtain drift. The Landau-Zener approach validates the intuitive picture of the diffusive dynamics as a quantum mechanical random walk driven by energy level crossings and elucidates the transition from localization to diffusion. This model could provide insight into solid-state systems involving electronic transport with a noisy bath. This approach, which is free from a constraint between $\tau$ and $T$, is also relevant to exciton diffusion in photosynthesis, where the broad range of time scales spanning many orders of magnitude requires a model with no constraints on noise correlation time \cite{goldilocks}.

\textit{Acknowledgments.-}  NP was supported by the Harvard College Research Program and Herchel Smith fellowship. AA thanks the Harvard Society of Fellows for support during the early stages of this work.

\bibliography{qdiffusion1}
\bibliographystyle{prsty}

\appendix

\section{Details of noise model}
\label{app:noise} 
 
In this section we determine the parameters $m,k,\eta$ of Eq. \eqref{eq:sde} in terms of the strength $W$, correlation time $\tau$, and mean velocity-squared $\langle v^2 \rangle$ of the noise we want to produce. The power spectrum, obtained by taking the square of the Fourier transform of $x$, is given by
\begin{equation}\label{eq:pwr}
S(\omega) = \frac{1}{m^2\omega^4+(\eta^2-2km)\omega^2+k^2}
\end{equation}
from which we determine the autocorrelation function using the Wiener-Khinchin theorem:

\begin{align}
G(t) &= \langle x(t') x(t'+t)\rangle = \frac{1}{2\pi} \int_{-\infty}^{\infty} S(\omega) e^{-i\omega t}\, d\omega\nonumber\\
 &= \frac{\frac1b e^{-b|t|} -\frac1a e^{-a|t|}}{2m^2(a^2-b^2)}
\end{align}
where 

\be
a = \frac{1}{2m} (\eta + \sqrt{\eta^2-4km}),\quad b = \frac{1}{2m} (\eta - \sqrt{\eta^2-4km}).
\ee

We take the correlation time to be 

\begin{equation}\tau = \max\left(a^{-1},b^{-1}\right) = \frac{2m}{\eta - \sqrt{\eta^2-4km}}.\end{equation}

For this to be valid, we require $\eta^2 \geq 4km$. Although this definition is adequate for our purposes, we note there is no unambiguous correlation time when $a \approx b$. We may also determine the variance of the noise to be $W^2 = G(0) = \frac{1}{2k\eta}$. 
\par

Next, applying the Fokker-Planck equation to the Langevin equation Eq. \eqref{eq:sde} (see, for instance, Ref. \cite{gardiner}) yields the stationary distribution
\begin{equation}\label{eq:pxv}
P(x,v) = \frac{\eta \sqrt{mk}}{\pi} \exp\left( - \eta k x^2 -\eta m v^2\right),
\end{equation}

giving $\langle v^2 \rangle \equiv c^2 \frac{W^2}{\tau^2} = \frac{1}{2m \eta}$, so that 

\be
c^2 = \frac{2}{1-\sqrt{1-4km/\eta^2}}-1.
\ee

This third parameter is necessary to describe a second-order process with confined position and velocity. We note that the requirement $\eta^2 \geq 4km$ is equivalent to $c\geq 1.$\par
Eq. \eqref{eq:subs} provides a mapping from the parameters $m,\eta,k$ of the Langevin equation to the noise-specific parameters $W,\tau, c$.

\section{Duration of Landau-Zener Transition}
\label{app:LZ}

Here we summarize the derivation of Eq. \eqref{eq:tLZ} given by Ref. \cite{mullen}. In this ``internal clock'' approach, we show that the the probability profile $P(t)$ is independent of $v$ and $T$ when time is measured in units of $t_{LZ}$.\par
We consider the time evolution of a state 

\begin{equation}
\psi(t) = A(t) |1\rangle + B(t) |2\rangle 
\end{equation}
under the Hamiltonian $H$ with $H_{11} =  vt/2$, $H_{22} = - vt/2$, and $H_{12} = T$. With initial condition $A(-\infty) = 1$, we want to find the time scale of the transition from $1$ to $e^{-2\pi T^2 / v}$ undergone by $P(t) \equiv |A(t)|^2$. \par
It is instructive to define the dimensionless parameters

\begin{equation}
g = \frac{T}{\sqrt{(v/2)}}, \qquad y = \sqrt{(v/2)} t.
\end{equation}
The Schr\"odinger equation for $\psi$ may be written equivalently as the coupled differential equations

\begin{align}
i \frac{d A}{dy} &= y A + g B\\\label{eq:A4}
i \frac{dB}{dy} &= -y B + g A.
\end{align}

For $g\ll 1$, we may expand $A\approx A_0 + g A_1 + g^2 A_2$ and $B\approx B_0 + g B_1 + g^2 B_2$ to solve this perturbatively. In particular, to $O(g^2)$ we must solve the equations

\begin{align}
i \frac{dA_0}{dy} &= yA_0\\
i \frac{dB_0}{dy} &= -yB_0 \\
ig \frac{dA_1}{dy} &= ygA_1 + gB_0\\
ig \frac{dB_1}{dy} &= -ygB_1 + gA_0\\
ig^2 \frac{dA_2}{dy} &= yg^2 A_2 + g^2 B_1
\end{align}
with initial conditions $A_i(-\infty) = \delta_{i0} ,  B_i(-\infty) = 0$ for $i =0,1,2...$, which sets $B_0=A_1=0$. This procedure yields the probability profile

\begin{equation}\label{eq:Py}
P(y) = 1- 2g^2 \left(\int\displaylimits_{-\infty}^{y} du \int\displaylimits_{-\infty}^u dv \cos(u^2 - v^2) \right) + O(g^4).
\end{equation}

Writing the term in parentheses as $F(y)$, we find that the rescaled probability profile $\frac{P(y)-P(\infty)}{1-P(\infty)} \approx -\frac{2}{\pi} F(y) + 1$ depends on $T,v$ and $t$ only through $y$. In this regime, therefore, $t_{LZ} \sim 1/\sqrt{v}$.  \par

For $g\gg 1$, we differentiate the Schr\"odinger equations for $A, B$ and substitute to obtain  

\begin{equation}
\frac{d^2A}{dy^2} +(g^2+y^2+i)A = 0.
\end{equation}

We solve this using the WKB method, treating the second term as the negative of our potential. This yields 

\begin{equation}
A \approx C \frac{e^{i \int_0^y \kappa(y')\,dy'}}{\sqrt{\kappa(y)}},\qquad \kappa = \sqrt{ g^2 + y^2 +i},
\end{equation}

and is valid when $\left|\kappa' \kappa^{-2}\right|\ll 1$ or more explicitly when $\left|y(g^2+y^2+i)^{-3/2}\right|\ll 1$. Since $g \gg 1$, this condition is satisfied for all $y$. Defining $x \equiv y/g$ and evaluating $P(y)$ gives us 

\begin{align}
P(y) &\approx \left|C \frac{e^{\frac{i}{2} y\kappa + (g^2+i) \log \frac{y+\kappa}{\sqrt{g^2+i}}}}{\sqrt{\kappa}}\right|^2\\
&\approx |C|^2 \frac{e^{-\log (x+ \sqrt{1+x^2})}}{g\sqrt{1+x^2}}.
\end{align} 

Imposing the boundary condition $P(x=-\infty) = 1$, we obtain the normalization $|C|^2 = g/2$. Finally, our probability profile

\begin{equation}\label{eq:P(x)}
P(x) \approx \frac{1}{2\left( \sqrt{1+x^2}\right)\left(x + \sqrt{1+x^2}\right)}
\end{equation}

depends on $t$ only through $x= \frac{vt}{2T}$, so we find that in this regime $t_{LZ} \sim T/v$. \par

\begin{figure}
\includegraphics[width=8.7cm]{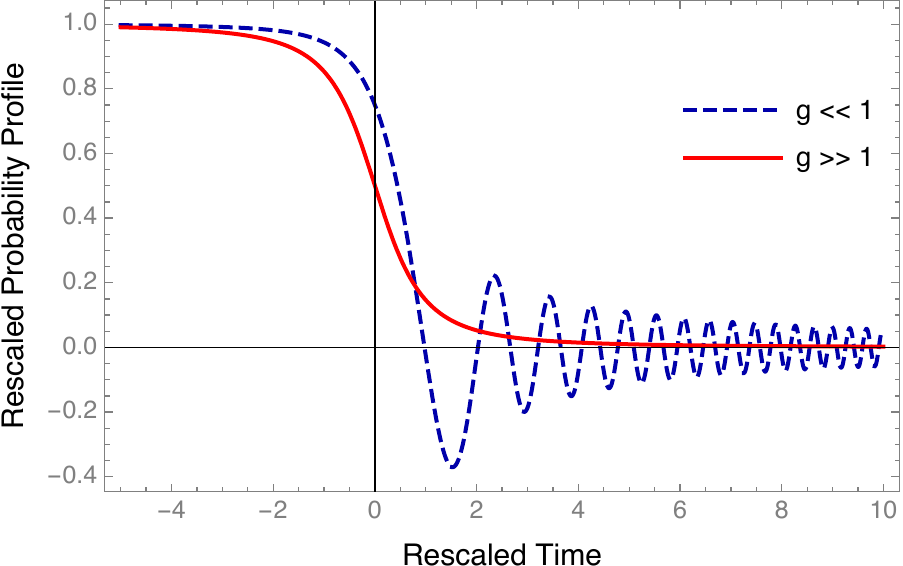}
\caption{Plots of the rescaled probability profiles for small and large $g$ (see Eqs. \eqref{eq:Py} and  \eqref{eq:P(x)}, respectively) as a function of rescaled, dimensionless time variables $y$ and $x$ respectively.} \label{fig:appA} 
\end{figure}

These results have been corroborated by further analytical and numerical studies, for instance Refs. \cite{vitanov,niu}, the latter of which also analyzes the crossover regime. Fig. \ref{fig:appA} shows plots of the rescaled probability profiles.

\section{Calculation of frequency of crossings}
\label{app:rice}

Here we calculate the frequency of crossings for two samples of the noise model defined by Eq. \eqref{eq:sde}, equivalent to the frequency $f$ of level zero crossings for a sample with twice the variance. We cite Rice's formula, first proved in Ref. \cite{rice}.\par

\textit{\textbf{Theorem:}
The expected number of level $u$-crossings per unit time of a stationary stochastic process $x(t)$ is}
\be
\label{eq:rice}
\mathbb{E}\{ C_u\} = \int_{-\infty}^{\infty} |v| P(u,v) \, dv
\ee

\textit{where $P(x,v)$ is the joint stationary distribution of $x(t)$ and its mean-square derivative $v(t)$.} \par

For a process with distribution given by 

\begin{equation}
\widetilde P(x,v) = \frac{\tau}{4\pi c W^2} \exp \left( - \frac{x^2}{4W^2} - \frac{v^2}{4c^2 W^2 / \tau^2}\right),
\end{equation} 

we obtain 

\begin{equation}\label{eq:C3}
\mathbb{E}\{C_u\} = \frac{c}{\pi \tau} \exp\left( - \frac{u^2}{4W^2}\right).
\end{equation}
Taking $u=0$ yields $f= \frac{c}{\pi \tau}$. In Section \ref{sec:E} we consider $u=E$, in which case Eq. \eqref{eq:C3} yields $f= \frac{c}{\pi\tau} e^{- \frac{E^2}{4W^2}}.$

\section{Addition of electric field}
\label{app:E}
Here we find the diffusion constant $D$ for a wavepacket evolving under the Hamiltonian in Eq. \eqref{eq:hamiltonian} with the addition of a time-independent electric potential $V_j = -j E$. This derivation is done in the regime $T\ll W$ and follows Ref. \cite{amir_PRE} closely. The Schr\"odinger equation for the site amplitudes $A_j$ is 

\begin{equation}\label{eq:schrodinger}
i \frac{dA_j}{dt} = T(A_{j+1} + A_{j-1}) +( x_j(t) +V_j)A_j
\end{equation}

where $x_j(t)$ is now a generic Gaussian noise.  It follows that the probabilities $P_j = |A_j|^2$ satisfy 

\begin{equation}\label{eq:dpjdt}
\frac{dP_j}{dt} = 2T \mbox{Im}[A_j^*(A_{j+1}+A_{j-1})].
\end{equation}

To zeroth order in $\frac{T}{W}$, Eq. \eqref{eq:schrodinger} is solved by $A^0_j = |A^0_j|e^{-i\int_0^t(x_j(t')+V_j)dt'} \equiv e^{-i(\phi_j(t)+V_jt)}$. To next order $A^1$ in $\frac{T}{W}$, we obtain the differential equation

\begin{equation}
i \frac{dA_j^1}{dt} -(x_j(t) +V_j) A_j^1 = T(A^0_{j+1} + A^0_{j-1}).
\end{equation}

Defining an integration factor $\mu_j = e^{i(\phi_j(t)+V_jt)}$, this can be rewritten in the form

\begin{equation}
\frac{d[A_j^1 \mu_j]}{dt} = -i \mu_j T(A_{j+1}^0 + A_{j-1}^0).
\end{equation}

Upon integration, we obtain

\begin{equation}\label{eq:Aj1}
A_j^1 = A_j^0 -i \frac{T}{\mu_j(t)} \int_0^t \mu_j(t') (A_{j+1}^0 + A_{j-1}^0)dt'.
\end{equation}

Using Eq. \eqref{eq:dpjdt} and taking the ensemble average gives us 

\begin{equation}\label{eq:ensembledpjdt}
\left\langle \frac{dP_j}{dt}\right\rangle \approx \langle 2T \mbox{Im}[(A_j^1)^*(A_{j+1}^1 + A_{j-1}^1)]\rangle.
\end{equation}

Since the noise at different sites is uncorrelated, upon plugging Eq. \eqref{eq:Aj1} into Eq. \eqref{eq:ensembledpjdt}, terms such as $(A_j^0)^*A_{j+1}^0$ will vanish in the ensemble average. There are four non-vanishing terms of similar form, one of which is $-2T^2 I$ with 

\begin{equation}
I = \left \langle \mbox{Im} \left[(A_j^0)^*\int_0^t \frac{i \mu_{j+1}(t')}{\mu_{j+1}(t)} A_j^0dt'\right]\right\rangle.
\end{equation}

Plugging in the explicit forms of $A_j^0$ and $\mu_j(t)$, we find 

\begin{equation}
I = \left \langle P_j \int_0^t e^{i[\phi_j(t)-\phi_j(t')]}e^{i[\phi_{j+1}(t)-\phi_{j+1}(t')]}  e^{i[V_{j+1} - V_j]t'} dt' \right\rangle
\end{equation}

We shall assume that the diffusion is a slower process than the dephasing, allowing us to use the separation of scales between the rate of change of the probabilities and the dephasing, as was the key point in Ref. \cite{amir_PRE}.
Defining the (site-independent) dephasing correlation function $C_{\phi}(t) \equiv \langle e^{-i\phi(t)}\rangle$, we then find that $I = P_j(t)Q$, with

\begin{equation}\label{eq:Q}
Q = \int_{-\infty}^{\infty} C_{\phi}^2 e^{-iE t} \, dt.
\end{equation}

From Eq. \eqref{eq:ensembledpjdt} we now obtain a classical diffusion equation with the diffusion constant given by 

\be
D = T^2 Q.
\ee

Hence, when $E$ is nonzero, $D$ is proportional to a Fourier component of $C_{\phi}^2$ rather than its integral. For noise whose correlation in time decays exponentially or as a Gaussian, the result is that the electric field leads to a \textit{smaller} diffusion constant. \par
 For $W\tau \gg1$ the correlation function takes the form $C_{\phi}(t) \approx e^{-W^2 t^2 /2}$ (Ref. \cite{amir_PRE}), for which we find

\begin{equation}
D=  \frac{\sqrt{\pi}T^2}{W} e^{-\frac{E^2}{4W^2}}.
 \end{equation}




\end{document}